\documentstyle[epsfig,rotating]{aipproc}

\begin{document}
\title{Strangeness, Charm, and Beauty: \\ Mesons with One Heavy Quark}

\author{Stephen Godfrey}
\address{Ottawa-Carleton Institute for Physics\thanks{This Research is 
supported by the Natural Sciences and Engineering Research Council of 
Canada}, \\
Department of Physics, Carleton University, Ottawa Canada K1S 5B6}

%\lefthead{LEFT head}
%\righthead{RIGHT head}
\maketitle

\begin{abstract}
I review the properties of mesons with one heavy quark.  I  
start by examining the predicted masses and widths in the context of 
the quark model and heavy quark effective theory. 
Some recent experimental results
are described and I conclude with comments on 
potentially interesting future experimental studies.
\end{abstract}

\section*{Introduction}

Mesons with one heavy quark constitute a beautiful laboratory to test our 
ideas of QCD.  
As the heavy quark's mass increases its motion decreases and the 
mesons' properties are increasingly governed by the dynamics of the 
light quark and approach a universal limit.  These mesons become 
the hydrogen atoms of hadron physics \cite{dgg76,rosner86}. 
In the heavy quark limit there are several rigorous 
treatments of QCD; the heavy quark effective theory (HQET) \cite{hqet}
and QCD on the lattice \cite{lattice}.
Mesons with one heavy quark provide a spectroscopy as rich as 
charmonium but because the relevant scales are governed by the light 
quark they probe different regimes.  As such they bridge the gap 
between heavy quarkonium and light hadrons providing an intermediate 
step on the way to studying the 
more complicated light quark sector in search 
of exotica like glueballs and hybrids.  There are many important 
topics that pertain to these mesons; spectroscopy, weak decays, and 
production in fragmentation \cite{falk96}.  
I do not pretend to review this entire field 
in the limited space available 
and will concentrate on topics with which I am most 
familiar.  In this vein I concentrate on spectroscopy and rely heavily
on the quark model which provides a useful guide to hadron 
spectroscopy and is a reasonable approximation to QCD.
I apologize in advance to the many workers in the field whose 
contributions I do not include.

\section*{The Constituent Quark Model}

We begin our description of mesons with one heavy quark using the 
constituent quark model as it is a good guide to hadron spectroscopy 
\cite{gi85,gk91}. 
However, it is important to 
remind ourselves that the quark model is not QCD but a very successful 
phenomenological model.  
In the non-relativistic limit the spin dependent pieces of the
interquark potential relevant to our discussion of
multiplet splittings are given by
\begin{equation}
H^{cont}_{q\bar{q}} = {{32 \pi} \over 9} 
{{ \alpha_s(r)} \over{m_{q}m_{\bar{q}} } } \vec{S}_q \cdot \vec{S}_{\bar{q}}
\delta^3 (\vec{r}) 
\end{equation}
which is the contact term and 
which splits the spin triplet-singlet degeneracy of S-wave mesons like  
the $D^* -D$ states.  The tensor interaction is given by 
\begin{equation}
H^{ten}_{q\bar{q}} = {4\over 3}
{ {\alpha_s(r)}\over {m_q m_{\bar{q}} } } {1\over{r^3}}
\left[{ 
{ { 3\vec{S}_q\cdot \vec{r} \; \vec{S}_{\bar{q}} \cdot \vec{r} } \over {r^2} } 
- \vec{S}_q \cdot \vec{S}_{\bar{q}} 
}\right] 
\end{equation}
and contributes to splittings of the $L\neq 0$ spin-triplet states and 
also gives rise to mixings of states with the same $J^{PC}$ but 
different $L$ values such as $^3S_1-^3D_1$.  There are two 
contributions to the spin-orbit interactions.
The first contribution is the Thomas precession term and is a purely 
relativistic effect which occurs when an object with spin moves in a 
central potential:
\begin{equation}
H^{s.o.(tp)}_{q\bar{q}} = - {1\over{2r}}
{ { \partial H_{q\bar{q}}^{conf} } \over{\partial r} }
\left({ { {\vec{S}_q} \over {m_q^2 } } 
+ { {\vec{S}_{\bar{q}} }\over {m_{\bar{q}}^2 } } }\right) \cdot \vec {L}
\end{equation}
The final spin-orbit term is the colour magnetic term which arises 
from the Lorentz vector nature of the short distance 
one-gluon-exchange:
\begin{equation}
H^{s.o.(cm)}_{q\bar{q}} = {4\over 3} {{\alpha_s(r)}\over {r^3}}
\left({ 
{ {\vec{S}_q} \over {m_q m_{\bar{q}} }} 
+ { {\vec{S}_{\bar{q}} }\over {m_q m_{\bar{q}}} } 
+ { {\vec{S}_q} \over {m_q^2 } } 
+ { {\vec{S}_{\bar{q}} }\over {m_{\bar{q}}^2 } } 
}\right) \cdot \vec {L} 
\end{equation}
Because of the origins of the different pieces, the colour-magnetic 
piece dominates at short distance while the Thomas precession term 
will become more important at larger $q\bar{q}$ separation. This results 
in an inversion of the ordering of the meson masses for larger orbital 
and radial excitations.  Studying where this inversion takes place 
tests this description of hadrons and can give us information 
about the underlying physics \cite{godfrey85,isgur97}.

For unequal mass quark and and anti-quark charge conjugation
parity is no longer a good quantum number and
the $^3L_L$ and $^1L_L$ states can mix via
the spin-orbit interaction or some other mechanism.
For example, the 
physical j=1 states are linear combinations of $^3P_1$ and $^1P_1$:
\begin{eqnarray}
Q_{low} & = & ^1P_1 \cos\theta + ^3P_1 \sin\theta \nonumber \\
Q_{high}& = & -^1P_1 \sin\theta + ^3P_1 \cos \theta  
\label{mix}
\end{eqnarray}
We will examine the physics of this mixing below.

The mass predictions of a particular quark model calculation 
\cite{gi85} are given 
in figure 1.

\begin{figure}[t!] % fig 1
%\vskip 11.5cm
\centerline{\epsfig{file=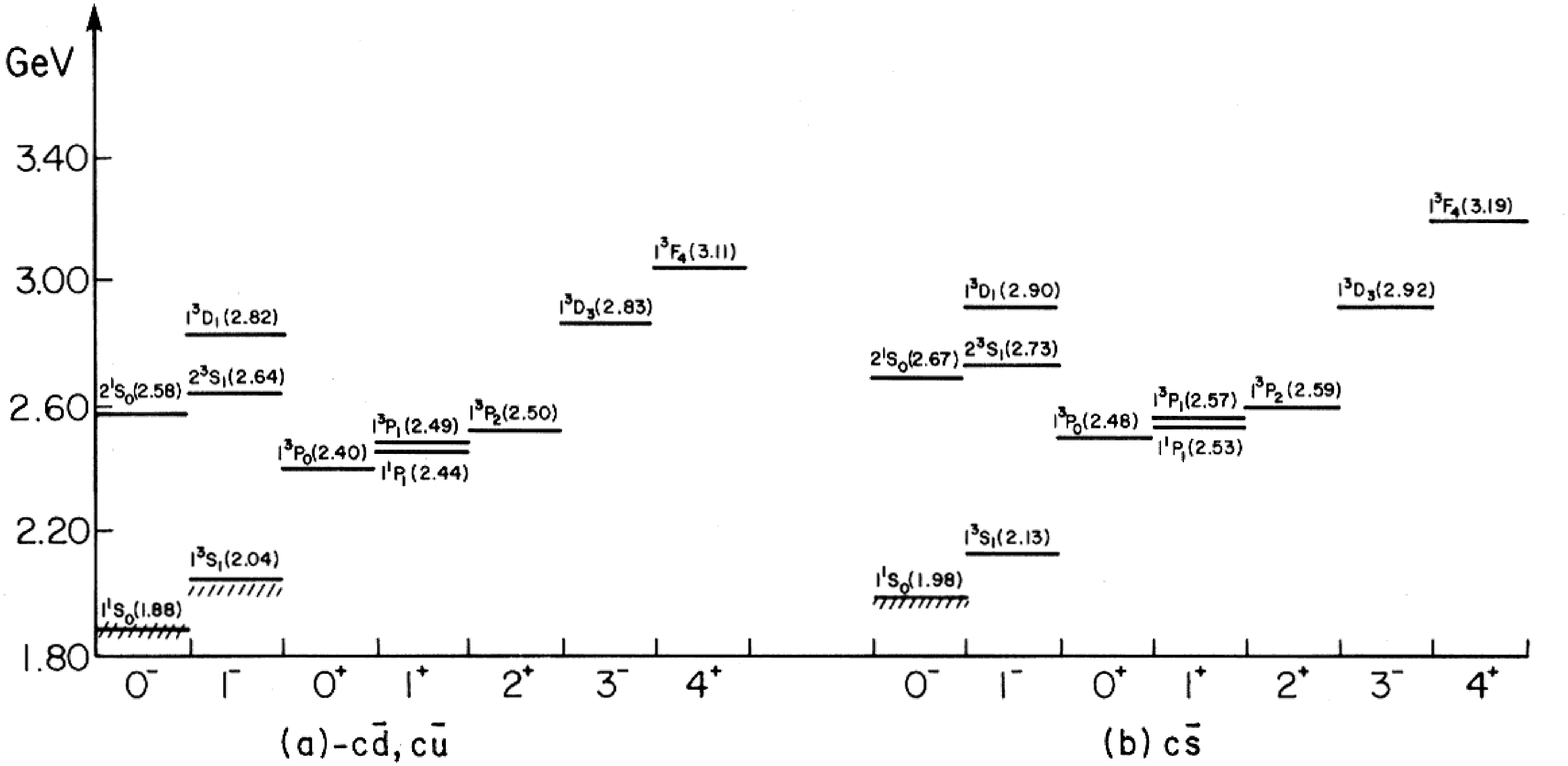,width=10.0cm,clip=}}
\centerline{\epsfig{file=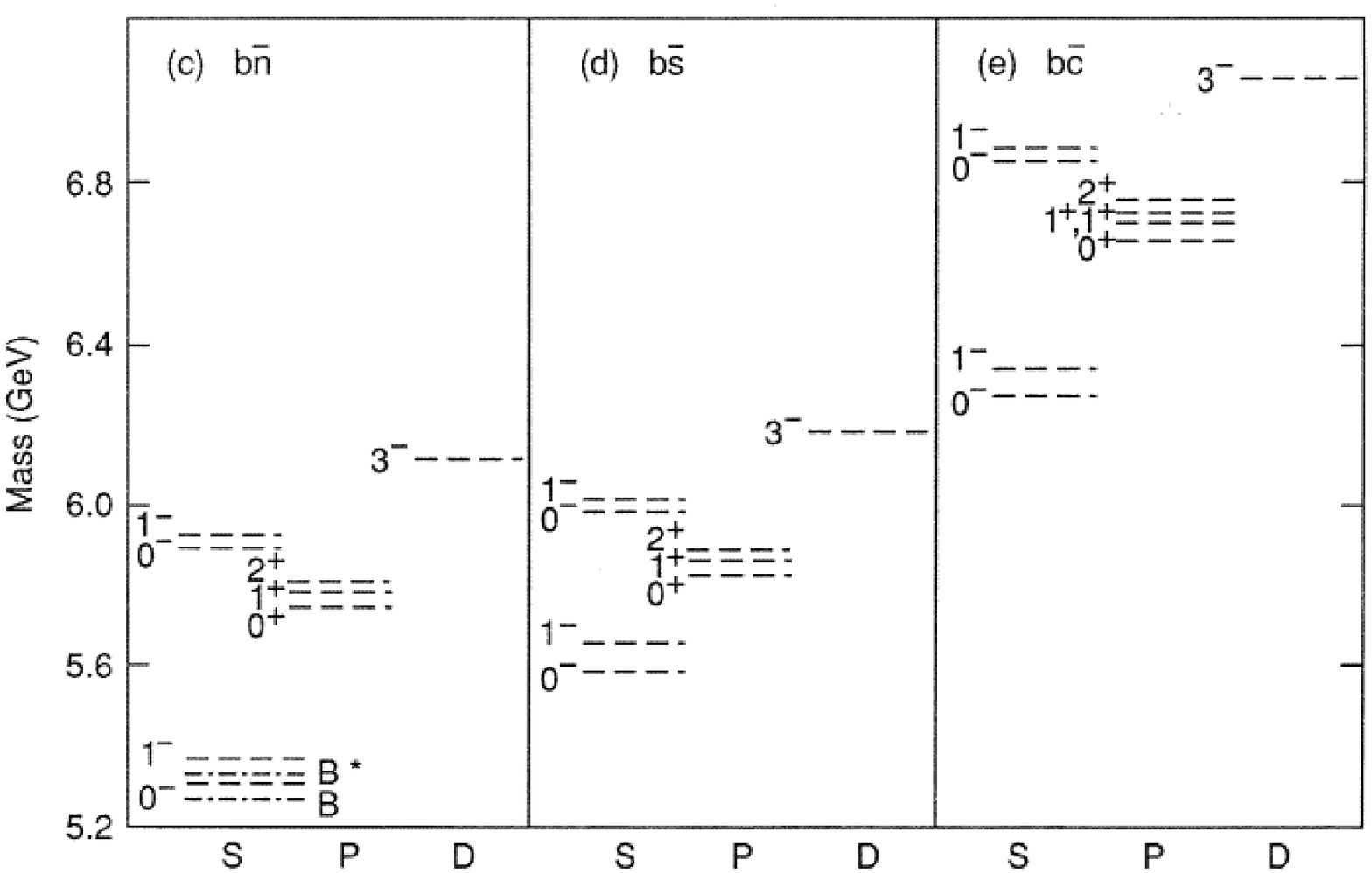,width=10.0cm,clip=}}
\vspace{10pt}
\caption[]{Mass predictions for mesons with one heavy quark.  From ref. 
\cite{gi85}. }
\label{fig1}
\end{figure}

Decay properties are sensitive to the internal structure of a state 
and therefore test how well a model describes the dynamics of the 
system. As a concrete example with experimental data I will concentrate 
on P-wave mesons.  
For such states OZI allowed decays can be described by two independent
amplitudes, S and D, which for the physical j=1 states 
are given by:
\begin{eqnarray}
A_S^{\bar{q}} (Q_{low} \to ^3S_1 \pi) &\sim & \sin(\theta+\theta_0 )S 
\nonumber \\
A_D^{\bar{q}} (Q_{low} \to ^3S_1 \pi) &\sim &\cos(\theta+\theta_0 )D 
\nonumber \\
A_S^{\bar{q}} (Q_{high} \to ^3S_1 \pi) &\sim &\cos(\theta+\theta_0 )S 
\nonumber \\
A_D^{\bar{q}} (Q_{high} \to ^3S_1 \pi) &\sim & -\sin(\theta+\theta_0 )D 
\nonumber 
\end{eqnarray}
where $\sin\theta_0 = \sqrt{1/3}$ and $\cos\theta_0 = \sqrt{2/3}$ 
(so $\theta_0 \simeq 35.3^o$) which provides a convenient 
reparametization of
angular momentum Clebsch-Gordan coefficients.  The $S$ and $D$ 
amplitudes are calculated using a specific decay model resulting in 
the general observations:
\begin{itemize}
\item Decay widths are sensitive to phase space with the partial 
widths $\propto q^{2L+1}$ where $L$ is the relative angular momentum of 
the final state mesons.
\item D-type amplitudes are relatively insensitive to the model.
\item S-type amplitudes are sensitive to the model but are in general 
quite large.
\item The partial widths are sensitive to the $^3P_1 - ^1P_1$ mixing angle
\end{itemize}

To gain some insights into the heavy quark limit we 
rewrite the spin-orbit terms in a more suitable form
\begin{eqnarray}
H_{s.o.} &=& {4\over 3} {{\alpha_s}\over{r^3}}
{{ \vec{S} \cdot \vec{L} } \over {m_q m_Q}}
+ {1\over 4} 
\left({ {4\over 3} {{\alpha_s}\over{r^3}} - {b \over r} } \right)
\left[ { \left( { {1\over{m_q^2}} + {1\over {m_Q^2}} }\right)
\; \vec{S}\cdot \vec{L} 
 + \left( { {1\over{m_q^2}} - {1\over {m_Q^2}} }\right) 
\; \vec{S}_-\cdot \vec{L} }\right] \nonumber \\
& = & H_{s.o.}^q \; \vec{S}_q\cdot \vec{L} + H_{s.o.}^Q 
\; \vec{S}_Q\cdot \vec{L}.\nonumber 
\end{eqnarray}
For the $P$-wave meson masses, in the heavy quark limit, the 
expectation values are
\begin{eqnarray}
M(^3P_2) & = & M_0 + \langle H^q_{s.o.} \rangle \nonumber \\
\left( \begin{array}{c}
M(^3P_1) \\ M(^1P_1) 
\end{array} \right) & = &
\left(  \begin{array}{cc}
 M_0 - \langle H^q_{s.o.} \rangle & -\sqrt{2} \langle H^q_{s.o.} \rangle \\
-\sqrt{2} \langle H^q_{s.o.} \rangle   & M_0  
\end{array} 
\right) 
\left( \begin{array}{c}  ^3P_1 \\ ^1P_1 \end{array} \right)
\nonumber \\
M(^3P_0) & = & M_0  -2  \langle H^q_{s.o.} \rangle \nonumber
\end{eqnarray}
Diagonalizing the $J=1$ mass matrix gives a $^3P_1 -^1P_1$ mixing 
angle given by 
$\sin\theta = \sqrt{2/3}$, $\cos\theta=\sqrt{1/3}$ and the 
$J=1$ masses are given by
$M_{low} = M_0 - 2\langle H_{s.o.}^q \rangle$ which is
degenerate with the $^3P_0$ state
and $M_{high} = M_0 + \langle H_{s.o.}^q \rangle$ which is 
degenerate with $^3P_2$ state.
Note that for $\langle H^q_{s.o.} \rangle < 0$ the ordering inverts so 
the ordering of the states can give information about the interquark 
potential.

 To measure the $^3P_1-^1P_1$ mixing angle examine the decays:
$$\Gamma (Q_{low}  \to ^3S_1 \pi ) \sim [S^2 \sin^2 (\theta 
+\theta_0 ) + D^2 \cos^2 (\theta + \theta_0 )] $$
$$\Gamma (Q_{high} \to ^3S_1 \pi ) \sim [S^2 \cos^2 (\theta 
+\theta_0 ) + D^2 \sin^2 (\theta + \theta_0 )] $$
which is illustrated in figure 2.
In the limit $m_Q \to \infty$ 
 the state degenerate with the $^3P_0$ is S-wave  and 
the state degenerate with the $^3P_2$ is D-wave.  
Thus, measuring the angular distribution of the decay products gives an 
estimate of the $^3P_1-^1P_1$ mixing angle and hence can potentially 
give information about the sign and magnitude of the mixing amplitudes 
which can be related to that of the spin-orbit potential.  In reality 
the physical situation is more complicated because terms proportional 
to $m_Q^{-1}$ are not neglible.  Using quark model predictions as 
input we find that these terms cannot be neglected even for $B$-mesons
(eg. the $B^*-B$ splitting).

\begin{figure}[t!] % fig 2
%\vskip 5.0cm
\centerline{\epsfig{file=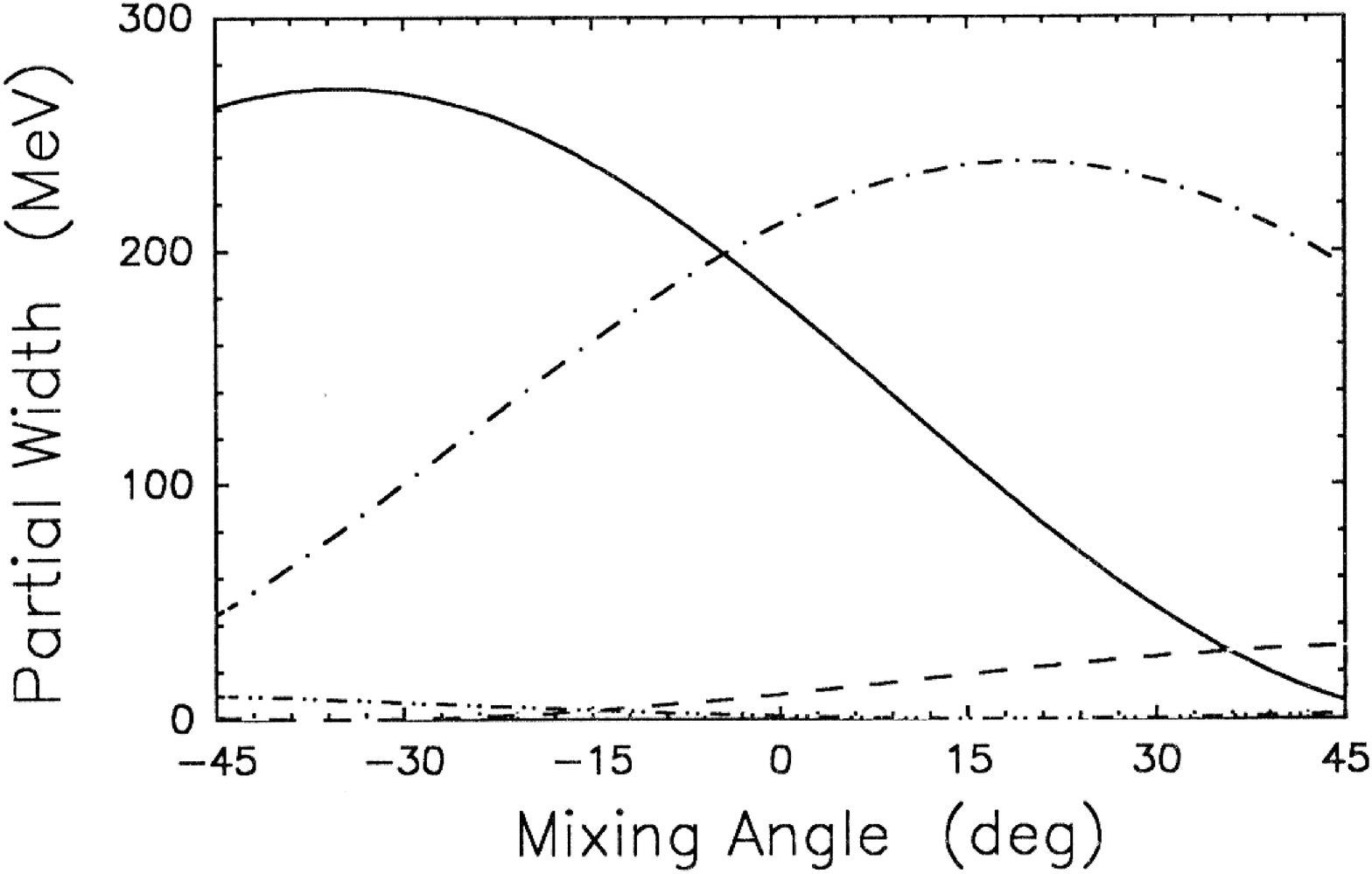,width=7.0cm,clip=}}
\vspace{10pt}
\caption[]{Partial decay widths $D_1 \to D^* \pi$ as a function of the 
$^3P_1-^1P_1$ mixing angle.  The solid line is for $Q_{high}$ S-wave,
the dashed line is for $Q_{high}$ D-wave, the dot-dashed line is
for $Q_{low}$ S-wave and the dot-dot-dashed line is for 
$Q_{low}$ D-wave.
\cite{gk91}. }
\label{fig2}
\end{figure}

An angular distribution analysis was performed by the CLEO 
collaboration \cite{cleo94}.  Defining the decay angles by 
${\rm D}^{(*)0}_J\to {\rm D}^{*+}\pi^- \to {\rm D}^0\pi^+ \pi^-$
with $\alpha$ the angle between the $\pi$'s, 
results in the angular distributions
\begin{equation}
\frac{dN}{d\cos\alpha}
\propto 
\left\{ {
\begin{array}{ll}
1-\cos^2\alpha & \mbox{for a $2^+$ decay} \\
1+3\cos^2\alpha & \mbox{for a D-wave $1^+$ decay}\\
1 & \mbox{for an S-wave $1^+$ decay}
\end{array} 
} \right. 
 \nonumber 
\end{equation}
The CLEO data and fits are shown in fig. 3. 
A fit to the data finds that 
pure $D$-wave is clearly a solution but a large $S$-wave 
content cannot be ruled out by the analysis.

\begin{figure}[h!] 
\begin{center}
%  \vspace{-3mm}
  \mbox{\epsfysize=3.8cm\epsffile[21 60 570 620]{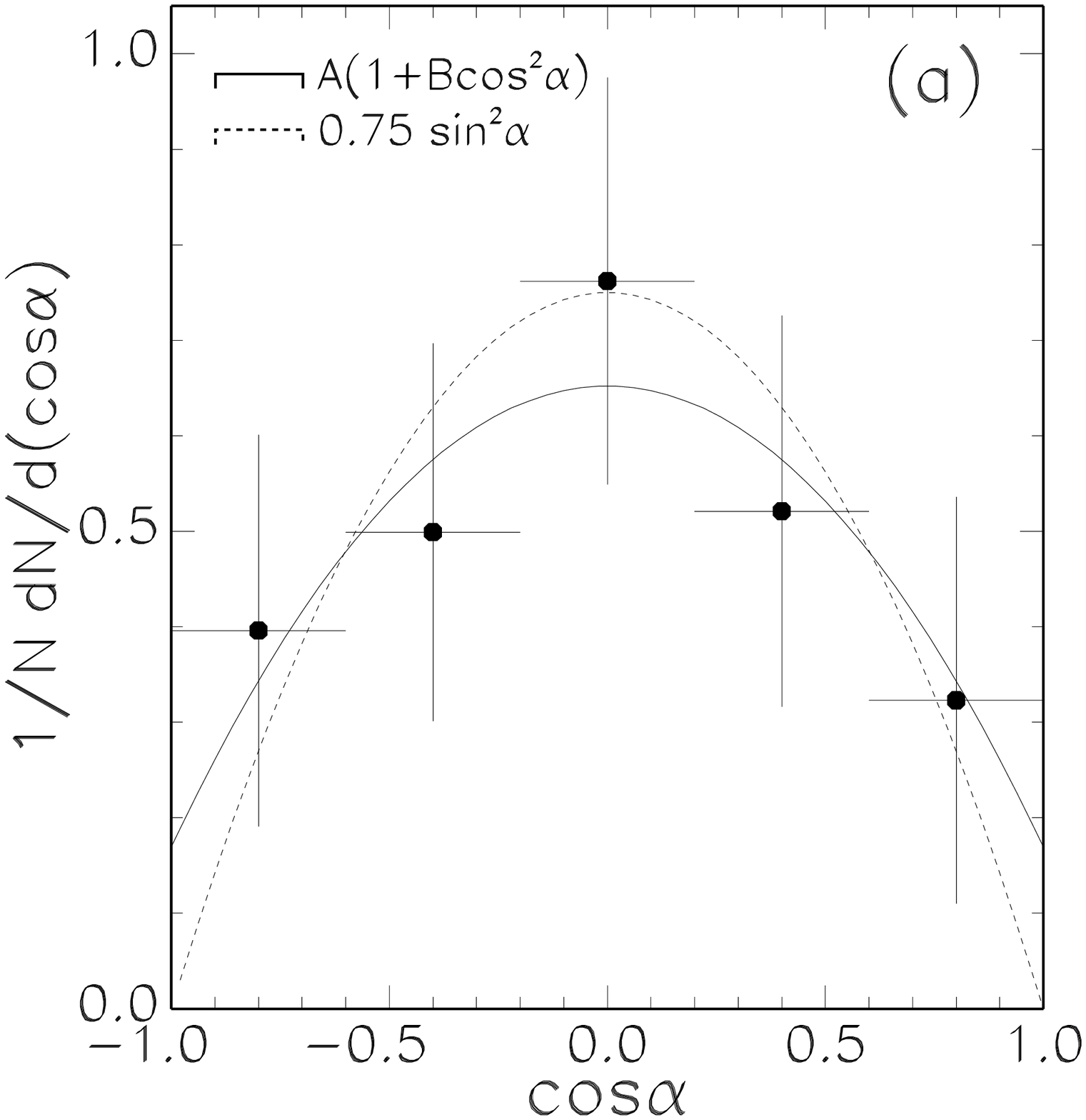} $\quad$
        \epsfysize=3.8cm\epsffile[21 60 570 620]{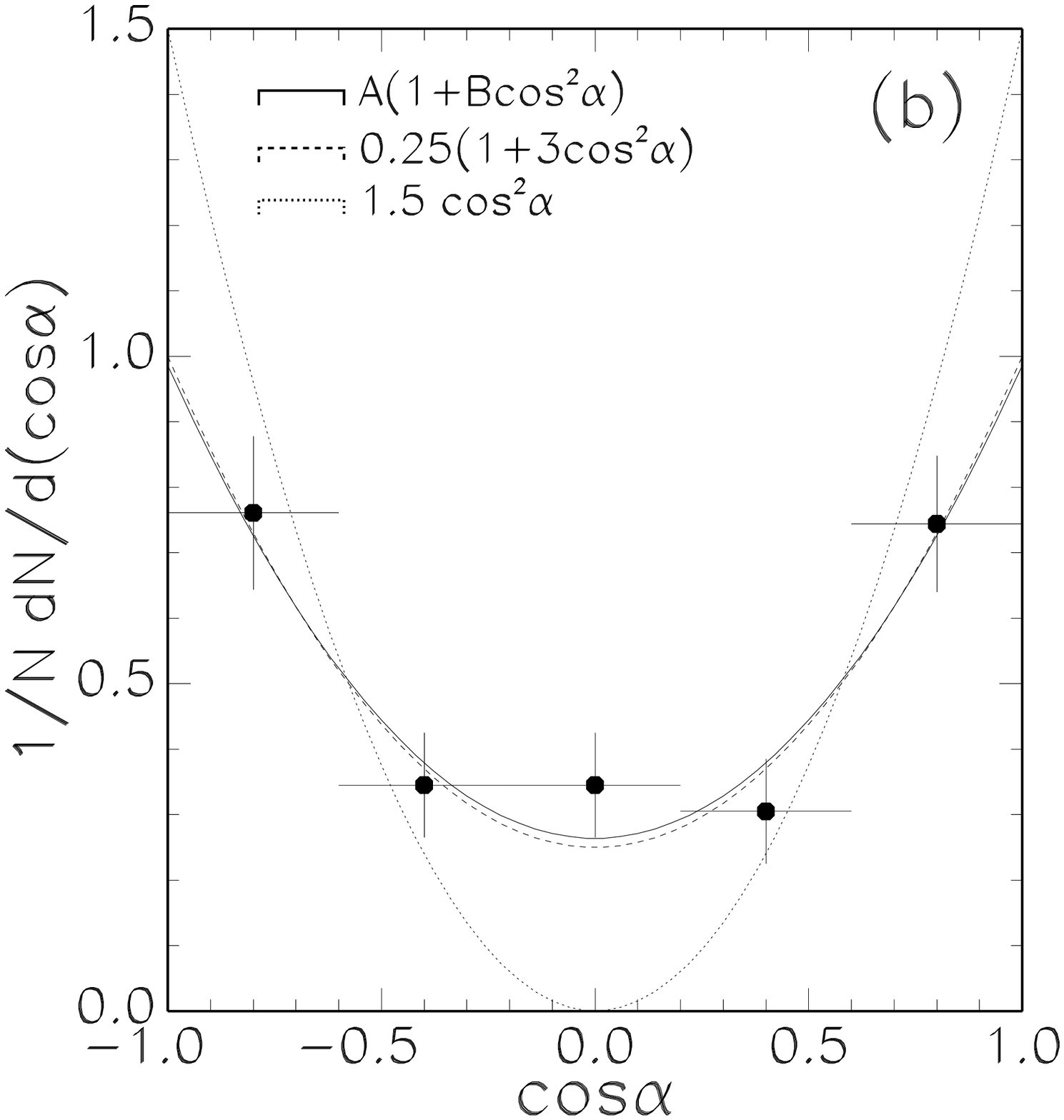}}
\end{center}
\vspace{10pt}
\caption[]{The normalized helicity angular distributions for 
(a) the $D_2^*$ and (b) the $D_1$ \cite{cleo94}. }
\label{fig3}
\end{figure}

\section*{The Heavy Quark Limit}

We started with the quark model since it provides a familiar 
framework to describe hadrons. In the heavy quark limit we found 
patterns in spectroscopy and decays which can be extended
to general principles \cite{hqet}.  
For a heavy quark where $m_Q >> \Lambda_{QCD} 
\sim 500$~MeV we can describe a hadron with a heavy quark as a bound 
state of a heavy quark with the lighter degrees of freedom 
characterized by $\Lambda_{QCD}$.  
Since for $m_Q >> \Lambda_{QCD}$ the heavy quark does not recoil due to 
exchange of light degrees of freedom and 
$Q$ acts as a static source of Chromoelectric field as far as 
the light degrees of freedom are concerned.  The 
light degrees of freedom are insensitive to $m_Q$.
Thus, heavy hadron spectroscopy differs from that for hadrons containing 
only light quarks because we may specify separately the spin quantum 
number of the light degrees of freedom and that of the heavy quark. ie 
$\vec{S}_Q$ and $\vec{j}_l = \vec{S}_q + \vec{L}$ are separately 
conserved so that each energy level in the excitation spectrum is 
composed of degenerate pairs of states 
$\vec{J} =\vec{j}_q + \vec{S}_Q = \vec{j}_q \pm 1/2$.  
In this language the ground state heavy mesons $M$ and $M^*$ have light 
degrees of freedom with $J_\ell^P = {1\over 2}^- $.
Combined with the spin of the heavy quark, 
$s_Q = {1\over 2}^- $, the two degenerate ground states have $J^P=0^-$ and 
$1^-$.  The next excitation is for the light quark in a P-wave with 
$J_\ell^P = {1\over 2}^+$ and $J_\ell^P = {3\over 2}^+$ which leads to 
degenerate doublets with $J^P = 0^+$ and $1^+$ and 
$J^P = 1^+$ and $2^+$ respectively.
This is exactly the result we obtained in the heavy quark limit for 
P-wave mesons in the constituent quark model.  It represents 
a new symmetry in the QCD spectrum in the heavy quark limit
and leads to relations between hadrons containing a single heavy quark
\cite{hqet,iw91}.
At this level HQET 
cannot tell us anything about the spectroscopy of the light quark 
system since in an expansion in powers of $1/m_Q$ the $D_2^*(2460) 
-D_1(2420)$ and $D^*(2010)-D(1870)$ splittings are $1/m_c$ effects.  

The power of the heavy quark symmetry is that it implies relations 
between the bottom and charmed meson systems since we can write
$$ M_H = m_Q +E_l.$$
Once the quark mass difference is known we can predict the entire 
bottom meson spectrum (with the experimental values in parenthesis): 
\begin{eqnarray}
B^* - B & = & 52 \; \mbox{MeV}  \; (46) \nonumber \\
B^*_s - B_s & = & 14 \; \mbox{MeV} \; (12) \nonumber \\
B^*_{s2} - B_{s1} & = & 13 \; \mbox{MeV} \;  (12) \nonumber \\
\bar{B}_{1} & = & 5789 \; \mbox{MeV} \; (5733) \nonumber \\
\bar{B}_{s1} & = & 5894 \; \mbox{MeV} \;  (5882) \nonumber
\end{eqnarray}
which is within the expected accuracy of these predictions.  
The importance of these results is that they follow rigorously 
from QCD in the heavy quark limit.

The two $D_1$'s
both have $J^P= 1^+$ and are only distinguished by $J_l$ 
which is a good quantum number in the limit $m_c\to \infty$.
Since strong decays are entirely transitions of the light quark 
degrees of freedom the decays from both members of a doublet with 
given $J_\ell$ to the members of another doublet with $J_\ell^P$ are 
all essentially a single process.  This leads to the simple prediction 
that the two excited states should have exactly the same widths.
$D_0^* \to D\pi$ and $D_1 \to D^* \pi$ decay via S-wave and can 
be quite broad and therefore difficult to identify experimentally.
In contrast
$D_1 \to D^* \pi$ and $D_2^* \to D^* \pi, \; D\pi$ proceed via D-wave 
and are much narrower.  
These states are identified as $D_1(2420)$ and $D_2(2460)$.  These are 
the same conclusions we obtained from the quark model.
More specifically, 
HQ symmetry predicts, after incorporating phase space and spin 
counting the predictions \cite{iw91}:
$$\frac{\Gamma(D_2^* \to D \pi)}{\Gamma(D_2^* \to D^* \pi)} =2.3 
\mbox{ vs } 2.2 \pm 0.96$$
$$\frac{\Gamma(D_1)}{\Gamma(D_2^*)}=0.3 \mbox{ vs } 0.71 $$
Falk and Mehen argue that this last 
discrepancy is due to terms of subleading  order
in the HQ expansion \cite{falk95}.

\section*{Recent Experimental Results}

The P-wave charmed mesons have been known for some time.  Recently the 
OPAL \cite{opal95}, ALEPH \cite{aleph}, and DELPHI \cite{delphi}
collaborations at LEP have reported the discovery of P-wave beauty mesons. 
The OPAL results are shown in fig 4.  Broad bumps are seen in $B\pi$ 
($M=5.68\pm 0.011$~GeV, $\Gamma=116\pm24$~MeV) and 
$BK$ ($M=5.853\pm 0.15$~GeV, $\Gamma=47\pm22$~MeV).  The $B\pi$ 
results are consistent with similar results by ALEPH and DELPHI. 
In both cases the 
widths are larger than the detector resolution of 40~MeV and the bumps 
are interpreted as superpositions of several states and/or decay 
modes.  In $B\pi$ the bump is assumed to be 
the $B_1$ and $B_2$ superimposed and in 
$BK$ the bump is assumed to be the $B_{2s}$ and $B_{1s}$ superimposed.

\begin{figure}[b!] 
\centerline{\epsfig{file=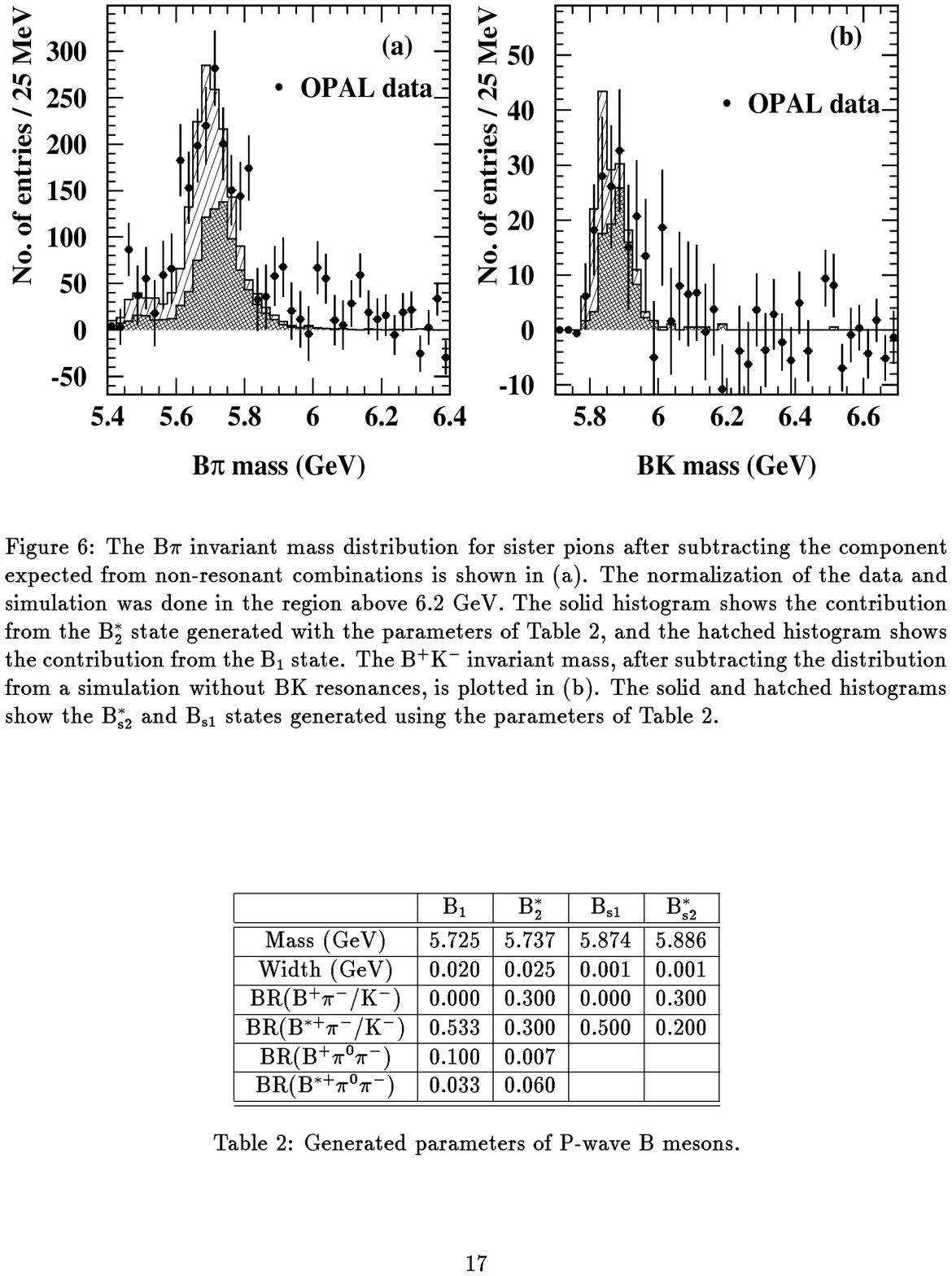,width=10.0cm,clip=}}
\vspace{10pt}
\caption[]{The $B\pi$ and $BK$ invariant mass distributions.  The solid 
histograms show Monte Carlo results for the $B_2^*$ and $B_{s2}^*$ 
states respectively and the hatched histograms show Monte Carlo 
results for  $B_1$ and $B_{s1}$ \cite{opal95}.}
\label{fig4}
\end{figure}

\section*{Strange Mesons}

Going from B mesons where the heavy quark limit is a reasonable 
approximation one might ask how well strange mesons are described by 
the heavy quark limit.  To answer this question we return to strong 
decays and also consider the weak couplings of the $K_1$'s.

The strong decays of the $K_1$ mesons can be described by 
equations \ref{mix} with similar expressions
for $K_1 \to [\rho K]_{S,D}$ and
$K_1 \to [\omega K]_{S,D}$ \cite{bgp96}.  
By comparing the measured partial widths to predictions of a model we 
can extract the $^3P_1 - ^1P_1$ mixing angle.  Comparing the HQ limit
\begin{eqnarray*}
& & \begin{array}{l}
j=1/2 \to K^* \pi \; \mbox{ in S-wave} \\
j=3/2 \to K^* \pi \; \mbox{ in D-wave} 
\end{array}
\end{eqnarray*}
to
\begin{eqnarray*}
& & \begin{array}{l}
K_1(1400) \to K^* \pi \;  \mbox{ is dominantly S-wave}   \\
K_1(1270) \to K^* \pi \; \mbox{ is a mixture of S and D}  
\end{array}
\end{eqnarray*}
we conclude that the $K_1(1400)$ is dominantly $j=1/2$ and the 
$K_1(1270)$ is dominantly $j=3/2$.    
A careful fit, using the various partial widths and the $D/S$ ratio,
finds that $\theta_K \simeq 45^o$  versus the HQ limit $\theta_K \simeq 
35^o$.

Weak decays can also be used to measure $\theta_K$ as $\Gamma (\tau 
\to K_1 \nu_\tau)$ varies with the mixing angle so that
$$ \frac{B[\tau \to \nu_\tau K_1(1270)] }{B[\tau \to \nu_\tau 
K_1(1400)] }$$ is a sensitive measure of $\theta_K$.
The axial meson decay constants through which these decays take place 
are defined by
\begin{eqnarray}
& & \langle 0 \vert \bar{q} \gamma^{\mu} (1-\gamma_5 ) q \; \vert
\; M(\vec{K},\lambda) \rangle = {i\over {(2\pi )^{3/2} } } f_{K_1} 
\epsilon^{\mu}(\vec{K},\lambda) 
\end{eqnarray}
Using the Mock-Meson approach\cite{cg90} the non-relativistic 
limit of the $f_{K_1}$ are given by:
\begin{eqnarray}
f_{K_1} (^3P_1)  & = & - \sqrt{12 M_{\widetilde{K_1}} } 
\left[ { {1\over m_q} + {1\over m_{\bar{q}}}  } \right] 
\left. \sqrt{3 \over{8\pi}}  {{\partial R_P (r)}\over {\partial r}} 
\right|_{r=0}  \nonumber \\
f_{K_1} (^1P_1) & = & \sqrt{6M_{\widetilde{K_1}} }
\left[ { {1\over m_q} - {1\over m_{\bar{q}}}  } \right]
\left. \sqrt{3 \over{8\pi}}  {{\partial R_P (r)}\over {\partial r}} 
\right|_{r=0} 
\label{fk}
\end{eqnarray}
Combining with the $^3P_1-^1P_1$ mixing gives:
\begin{eqnarray}
f_{K_{low}}  & = & - A
\left[\left( { {1\over m_u} - {1\over m_s}  } \right) \cos\theta_K
-\sqrt{2}  \left( { {1\over m_u} + {1\over m_s}  } \right) \sin\theta_K
\right] \nonumber \\
f_{K_{high}}  & = & + A
\left[\left( { {1\over m_u} - {1\over m_s}  } \right) \sin\theta_K
+\sqrt{2}  \left( { {1\over m_u} + {1\over m_s}  } \right) \cos\theta_K
\right]
\label{nrfk}
\end{eqnarray}
where the definition of $A$ follows by comparing eqn. \ref{fk} to eqn. 
\ref{nrfk}.
The $K_b$ coupling goes like the SU(3) breaking $(m_s-m_d)$ so that
in the SU(3) limit the $K_b$ $(^1P_1)$  coupling goes to zero and 
only $K_a$ $(^3P_1)$ couples to the weak current.

To extract $\theta_K$ from the weak decay constants we consider the 
measurements 
of the TPC/$2\gamma$ collaboration \cite{tpc}
\begin{eqnarray*}
BR (\tau \to \nu K_1(1270)) & = & (0.41^{+.41}_{-.35} )\times 10^{-2} \\
BR (\tau \to \nu K_1(1400)) & = & (0.76^{+.40}_{-.33} )\times 10^{-2} \\
BR (\tau \to \nu K_1) & = & (1.17^{+.41}_{-.37} )\times 10^{-2} .
\end{eqnarray*}
We use the ratio
%\begin{equation}
$$R  = \frac{B[\tau \to \nu K_1 (1270)]}{B[\tau \to \nu K_1 (1400)]} 
 =  1.83 \left| {{\sin\theta_K - \delta \cos\theta_K}
\over {\cos\theta_K + \delta \sin\theta_K}} \right|^2 $$
%\end{equation}
where 1.83 is a phase space factor and 
$\delta$ is an $SU(3)$ breaking factor given by
%\begin{equation}
$\delta = {1\over \sqrt{2}} \left( {{m_s-m_u}\over {m_s + m_u}} 
\right)$ since theoretical uncertainties mainly cancel.
%\end{equation}
From a fit of $\theta_K$ to $R$ one finds that
$-30^o \lesssim \theta_K \lesssim 50^o$ at 68 \% C.L..
However, recent data by ALEPH, not 
included in the fit,  supports a relatively large BR($\tau \to K_1(1270) \nu$) from 
the $\rho K$ decay mode \cite{lafferty}.  Clearly better data is 
needed.

\section*{$D_r^*$ and $B^*_r$}

The final topic is the recent evidence for radial 
excitations of the $D_r^*$ and $B_r^*$ by the DELPHI collaboration.  
From the invariant mass distribution $M(D^*\pi\pi)$ 
Delphi obtains the mass measurement of $M(D_r^*) =2637\pm 2 \pm 6$~MeV 
which is in good agreement with the quark model prediction \cite{gi85}.
Likewise, from $Q(B^{(*)}\pi^+\pi^-)$ 
they obtain 
$M_{B^*_r}=5904\pm 4 \pm10$ which is also in good agreement with the 
quark model.
Although these results are preliminary they are promising and
together with the $B^{**}$ observations show the potential of high 
energy colliders for contributing to our understanding of hadron 
spectroscopy.

\section*{Summary}

Excited hadrons with one heavy quark exhibit as rich a phenomenology as
heavy quarkonium and light quark mesons.  They provide a bridge 
between the two regimes.  In the heavy quark limit we can obtain 
rigorous results from QCD making these states a good place to test models.  
We found that the Quark Model continues to be a reliable guide to 
hadron spectroscopy.
I only touched one aspect here, spectroscopy and decays.  There are
many other topics of importance,
such as semileptonic decays and production in fragmentation, which 
were omitted for lack of space.  
The continued study of mesons with one heavy quark
will pay large physics dividends and should be pursued.  In particular;
better measurements of $\tau \to K_1 \nu$ and other strange final 
states
would be useful for the 
study of meson dynamics, 
more $B$ results would add to our understanding, and if at all possible
angular distributions of $B^{**}$ are important tests of our 
understanding.

\end{document}